\documentclass[journal,comsoc] {IEEEtran}
\usepackage{amsmath,amsfonts,amssymb}
\usepackage{graphicx}
\usepackage{xcolor}
\usepackage{subfig}
\usepackage{makecell}
\usepackage{multirow}
\usepackage{tikz}
\usepackage{mathrsfs}
\usetikzlibrary{calc,arrows,shapes,chains,positioning,shapes.geometric}
\usepackage{enumitem}
\usepackage{algpseudocode}
\usepackage{algorithm}

\usepackage{amsthm,bm}
\usepackage{footnote}
\usepackage{pgfplots}
\pgfplotsset{width=10cm,compat=1.9}
\pgfplotsset{every axis legend/.style={
    cells={anchor=center},
    inner xsep=3pt,inner ysep=2pt,
    nodes={inner sep=2pt,text depth=0.1em},
    anchor=north east,
    shape=rectangle,
    fill=white,draw=green,font=\footnotesize,
}}
\usepackage{booktabs}
\usepackage{fancyhdr}
\usepackage{svg}

\fancypagestyle{plain}{
    \fancyhf{}
    \fancyfoot[C]{\thepage}
    
}

\usepackage{hyperref}
\hypersetup{
    anchorcolor=black, 
    bookmarksnumbered=true,
    bookmarksopen=true,
    bookmarksopenlevel=1,
    colorlinks=true,
    linkcolor=blue, 
    citecolor=red,
    urlcolor=magenta
}

\begin{document}
\thispagestyle{fancy}
\bibliographystyle{IEEEtran}
\title{Iterative decoding of short BCH codes and its post-processing}
\author{Guangwen Li, Xiao Yu
\thanks{G.Li is with  Shandong Technology and Business University, Yantai, China}
\thanks{X.Yu is with  Binzhou Medical University, Yantai, China.}
}

\maketitle

\begin{abstract}
Effective iterative decoding of short BCH codes faces two primary challenges: identifying an appropriate parity-check matrix and accelerating decoder convergence. To address these issues, we propose a systematic scheme to derive an optimized parity-check matrix through a heuristic approach. This involves a series of binary sum and row shift operations, resulting in a low-density, quasi-regular column weight distribution with a reduced number of shortest cycles in the underlying redundant Tanner graph. For the revised normalized min-sum decoder, we concurrently integrate three types of random permutations into the alternated messages across iterations, leading to significantly faster convergence compared to existing methods. Furthermore, by utilizing the iterative trajectories of failed normalized min-sum decoding, we enhance the reliability measurement of codeword bits with the assistance of a neural network model from prior work, which accommodates more failures for the post-processing of ordered statistics decoding. Additionally, we report the types of undetected errors for the design of iterative decoders for short BCH codes, which potentially challenge efforts to approach the maximum likelihood limit. Extensive simulations demonstrate that the proposed hybrid framework achieves an attractive balance between performance, latency, and complexity.
\end{abstract}


\thispagestyle{fancy}
\section{Introduction}
\label{intro_sec}
\IEEEPARstart{F}{or} low-density parity-check (LDPC) codes \cite{gallager62}, belief propagation (BP) \cite{mackay96} and its variants dominate decoding due to their asymptotic maximum-likelihood (ML) performance and high data throughput, enabled by parallelizable implementations. However, it remains challenging for BP variants to perform effectively on classical linear block codes with high-density parity-check (HDPC) matrices, such as Bose–Chaudhuri–Hocquenghem (BCH) and Reed–Solomon (RS) codes. The inherent abundance of short cycles in their underlying Tanner graphs (TGs) disrupts the assumed message independence necessary for any BP variant to achieve optimal decoding. This challenge has spurred significant interest in designing effective decoding schemes for HDPC codes within the coding community.

The importance of fully exploiting code structure to facilitate decoding has long been recognized. Jiang et al. \cite{jiang2004iterative} proposed stochastic shifting mechanisms and varied damping coefficients for the updated log-likelihood ratios (LLRs) of codeword bits. In \cite{jiang2006iterative}, they adapted the parity-check matrix $\mathbf{H}$ based on the consecutively updated reliability measurements of codeword bits across iterations, aiming to prevent BP from becoming trapped in pseudo-equilibrium points that hinder convergence to valid codewords. Halford et al. \cite{halford2006random} introduced random redundant decoding (RRD) by expanding cyclic permutations into the automorphism group of the relevant code, demonstrating improved bit error rate (BER) performance with a cycle-reduced $\mathbf{H}$. The modified RRD (mRRD) algorithm \cite{dimnik2009improved} subsequently fixed the damping factor and utilized multiple decoders in parallel, each assigned a permutation of the same base $\mathbf{H}$. The resultant complexity increase of \cite{dimnik2009improved} is expected to be lower than that of \cite{hehn2010multiple}, which employs multiple parity-check matrices composed of cyclic shifts of various minimum weight codewords from the dual code. Ismail et al. \cite{ismail2015efficient} presented a permuted BP (PBP) scheme that executes a randomly chosen automorphism group element for the messages in each BP iteration. However, the performance gains achieved by the aforementioned schemes come at the expense of multiple nested loops of BP decoding or serial Gaussian elimination operations on the $\mathbf{H}$, severely weakening overall throughput. In \cite{babalola2019generalized}, a generalized parity-check transformation (GPT) scheme was developed to update bit reliability using on-the-fly syndrome. However, this method suffers from poor BER performance and lacks parallelizability.

Given the significant impact of different parity-check matrices on BP performance, Lucas et al. \cite{lucas1998iterative} favored the $\mathbf{H}$ composed of minimum weight codewords from the dual code due to its relatively low density of non-zero elements. Yedidia et al. \cite{yedidia2002generating} proposed replacing the standard $\mathbf{H}$ with an expanded form that supplements auxiliary "bits" to reduce row weight and short cycles. Notably, \cite{kou01} validated that adding redundant rows to the $\mathbf{H}$ may potentially improve decoding, despite the risk of introducing additional short cycles.

Inspired by the success of deep learning techniques in other fields, Nachmani et al. \cite{nachmani18} proposed adapting existing BP variants by weighting messages flowing along the edges of the TG. Lian et al. \cite{lian2019learned} demonstrated that shared parameters and parameter evaluation for differing signal-to-noise ratios (SNRs) can alleviate the complexity of \cite{nachmani18} without sacrificing performance. However, the improvements in frame error rate (FER) or BER remain too modest to justify the accompanying increase in computational complexity across a wide range of waterfall SNR regions.

On the other hand, universal ordered statistics decoding (OSD) \cite{Fossorier1995} variants can fulfill decoding tasks while omitting specific code structures. Recently, Bossert et al. \cite{bossert2022hard} proposed utilizing formation set decoding as an OSD variant to tackle BCH code decoding, demonstrating competitive FER performance against complexity. However, since OSD essentially operates in serial mode, it is more suited for auxiliary roles rather than handling the primary decoding load in latency-sensitive scenarios.

To pursue the effective application of iterative BP decoding for short BCH codes while maintaining manageable computational complexity and achieving high decoding throughput with lower latency, we propose a concatenation strategy in which OSD decoding is invoked only when the normalized min-sum (NMS, the simplest BP variant) decoder fails. This hybrid framework, presented in our prior work \cite{li2024boosting} for LDPC codes, is now extended to HDPC codes with specific cyclic structures in this paper. The main contributions are outlined as follows:
\begin{itemize}
\item[*] An optimization method is proposed for deriving a $\mathbf{H}$ with lower density, a reduced number of short cycles, and evenly distributed column weights—all vital components for BP decoding.
\item[*] A revised NMS is proposed that aggregates three types of permutations from the automorphism group concurrently per iteration to significantly expedite BP decoding.
\item[*] The benefits of decoding information aggregation (DIA) proposed in \cite{li2024boosting} for LDPC codes are found to remain valid for HDPC codes.
\item[*] We initially expose that undetected FER may hinder the efforts of any iterative decoder design to approach ML limit for short BCH codes in practical applications.
\end{itemize}

The remainder of the paper is structured as follows: Section \ref{preliminary} reviews the preliminaries of NMS and mRRD variants. Section \ref{motivations} elaborates on deriving an improved $\mathbf{H}$, the revised NMS, and its collaboration with OSD. Section \ref{simulations} presents experimental findings and complexity analysis for short BCH codes. Concluding remarks are provided in Section \ref{conclusion}.

\section{Preliminaries}
\label{preliminary}
Assume a binary message row vector $\mathbf{m} = [m_i]_1^K$ is encoded into a codeword $\mathbf{c} = [c_i]_1^N$ via $\mathbf{c} = \mathbf{mG}$ in Galois field GF(2), where $K$ and $N$ are the lengths of the message and codeword, respectively, and $\mathbf{G}$ is the generator matrix. Each bit $c_i$ is mapped to an antipodal symbol using binary phase shift keying (BPSK) modulation, given by $s_i = 1 - 2c_i$. Due to additive white Gaussian noise (AWGN) $n_i$ with zero mean and variance $\sigma^2$, the received sequence at the channel output is $\mathbf{y} = [y_i]_1^N$, where $y_i = s_i + n_i$. This sequence is then sent to the decoder for codeword estimation.

The log-likelihood ratio (LLR) for the $i$-th bit is calculated as follows:
\begin{equation}
{l_i} = \log \left( \frac{{p(y_i|{c_i} = 0)}}{{p(y_i|{c_i} = 1)}} \right) = \frac{{2y_i}}{{\sigma^2}}.
\end{equation}
A larger magnitude of $y_i$ implies greater confidence in the hard decision for the $i$-th bit, a feature utilized by most OSD variants in decoding. Unlike standard BP decoding, the normalized min-sum (NMS) variant benefits from channel invariance \cite{lugosch18-1xVPf}, allowing $\sigma^2$ to be any constant, evaluated as 2 hereafter.

\subsection{Original NMS}
The TG of a code is decided totally by its $\mathbf{H}$ of dimensions $(N-K)\times N$. Variable nodes $v_i, i=1,2,\ldots,N$ exchange messages with check nodes $c_j, j=1,2,\ldots,(N-K)$ if there are nonzero entries in the $j$-th row and $i$-th column of $\mathbf{H}$.

In NMS decoding within the LLR domain, assuming a flooding schedule for message passing, the message from $v_i$ to $c_j$ at the $t$-th iteration ($t=1,2,\ldots,I$) is given by
\begin{equation}
x_{v_i \to c_j}^{(t)}  = {l_i} + \sum\limits_{\substack{c_p \to v_i\\p \in \mathcal{C}(i)\backslash j}} {x_{c_p \to v_i}^{(t - 1)}}
\label{eq_v2c}
\end{equation}
while the message from $c_j$ to $v_i$ is 
\begin{equation}
x_{c_j \to v_i}^{(t)} = \alpha \cdot \prod\limits_{\substack{v_q \to c_j\\ q \in \mathcal{V}(j)\backslash i}} sgn\left( x_{v_q \to c_j}^{(t)} \right) \cdot \min_{\substack{v_q \to c_j\\ q \in \mathcal{V}(j)\backslash i}} \left| x_{v_q \to c_j}^{(t)} \right|
\label{eq_c2v}
\end{equation}
where $\alpha$ is the normalization factor, $\mathcal{C}(i)\backslash j$ denotes all neighboring check nodes of $v_i$ except $c_j$, and $\mathcal{V}(j)\backslash i$ denotes all neighboring variable nodes of $c_j$ except $v_i$. All $x_{c_p \to v_i}^{(0)}$ terms are initialized to zero.

Messages update iteratively via \eqref{eq_v2c} and \eqref{eq_c2v} along the edges of the TG until the maximum iteration count $I$ is reached or a stopping criterion is met. Meanwhile, a posteriori LLRs of codeword bits at the $t$-th iteration are evaluated as 
\begin{equation}
x_{i}^{(t)} = {l_i} + \sum\limits_{\substack{c_p \to v_i\\p \in \mathcal{C}(i)}} {x_{{c_p} \to {v_i}}^{(t - 1)}}
\label{eq_bit_decision}
\end{equation}
from which a tentative hard decision $\hat{\mathbf{c}} = [\hat{c}_i^{(t)}]_1^N$ is derived and checked for the stopping criterion:
\begin{equation}
\label{termination_criterion}
\mathbf{H}\hat{\mathbf{c}}^{\top}=\mathbf{0}.   
\end{equation}

\subsection{mRRD and PBP}
It is recognized that $\rho(\mathbf{H})\hat{\mathbf{c}}^{\top}$ is equivalent to $\mathbf{H}\rho^{-1}(\hat{\mathbf{c}})^{\top}$, where $\rho(\cdot)$ is a specific permutation of codeword bits. Hence, for cyclic BCH codes, any decoder applying allowed permutations to $\mathbf{H}$ implies the inverse permutations on its input data.

As shown in Algorithm~\ref{alg::mrrd}, the mRRD deploys $I_3$ decoders in parallel to achieve diversity gain. Random permutations are applied to $\mathbf{y}$ after every $I_1$ iterations of BP decoding for a total of $I_2$ rounds. The codeword in the candidate list with the minimum Euclidean distance to $\mathbf{y}$ is selected as the final estimate. In comparison, the PBP scheme applies one random permutation per NMS iteration and exits as soon as a valid codeword is found to expedite decoding.

\begin{algorithm}[ht]
\caption{\hspace{1cm}mRRD decoder \cite{dimnik2009improved}}
\label{alg::mrrd}
\begin{algorithmic}[1]
\Require
$\mathbf{y}$, $\mathbf{H}$, automorphism group $\mathbf{P_g}$, $S = \varnothing $
\Ensure
Optimal codeword estimate $\hat{\mathbf{c}}$ for $\mathbf{y}$.  
\State
\textbf{for} {$i = 1, 2, \cdots,I_3$}
\State
\hspace{0.3cm}Draw a random element $\rho$ from $\mathbf{P_g}$; $\mathbf{y}\gets\rho(\mathbf{y})$; $\Theta\gets\rho$
\State
\hspace{0.3cm}\textbf{for} {$j = 1, 2, \cdots,I_2$}
\State
\hspace{0.6cm}Perform $I_1$ iterations of BP with soft outputs $\mathbf{y_s}$; update $\mathbf{y}\gets\mathbf{y}+\mathbf{y_s}$; obtain $\mathbf{\hat{c}}_i$ as the hard decision of $\mathbf{y}$.
\State
\hspace{0.6cm}\textbf{if }{$\mathbf{H}\hat{\mathbf{c}}^{\top}=\mathbf{0}$} 
{ $S\gets S\cup i$; apply $\Theta^{-1}$ to $\mathbf{y}$ and $\hat{\mathbf{c}}_i$; \textbf{break}}
\State
\hspace{0.6cm}\textbf{else}
{ Draw a new $\rho$ from $\mathbf{P_g}$; $\mathbf{y}\gets\rho(\mathbf{y})$; $\Theta\gets\rho\cdot\Theta$}
\State
\textbf{if}{ $S = \varnothing $} {  $S\gets \{1,2,\cdots,I_3\}$} 
\State
${\mathbf{\hat{c}}=\arg\min_{i \in S}\sum_{k=1}^{N}\left|y_k-s_k \right|^2}$ where $\mathbf{s}=1-2\hat{\mathbf{c}}_{i}$.

\end{algorithmic}
\end{algorithm}

\section{Motivations}
\label{motivations}
\subsection{Derivation of a New Parity-Check Matrix}
Our strategy is to derive an improved parity-check matrix $\mathbf{H}_s$ by applying a series of row operations and introducing redundancy to the standard $\mathbf{H}$. The goals are to achieve lower density, fewer short cycles, and a balanced distribution of row and column weights—all conducive to effective NMS decoding.

Assume $S_f = S_g = \varnothing$, where $||\cdot||$ denotes the Hamming weight. The standard form $\mathbf{H}_{(N-K) \times N}$ evolves as follows:
\begin{enumerate}
    \item Convert $\mathbf{H}$ into its row echelon form $\mathbf{H}_r$.
    \item Reduce $\mathbf{H}_r$'s density by binary sums on rows $\mathbf{r}_i$ and $\mathbf{r}_j$.
    
    \textbf{for} $i=1,2,\ldots,N-K$ 
    
\hspace*{0.3cm}$w_g = ||\mathbf{r}_i||; \; S_f \gets \left\{ \mathbf{r}_i \right\}$

\hspace*{0.3cm}\textbf{for} $j=1,2,\ldots,N-K, \; j\neq i$

\hspace*{1.0cm}$\mathbf{t} = \mathbf{r}_i \oplus \mathbf{r}_j$

\hspace*{1.0cm}\textbf{if} $||\mathbf{t}|| < w_g$  
  $w_g \gets ||\mathbf{t}||; \; S_f \gets \left\{ \mathbf{t} \right\}$ 

\hspace*{1.0cm}\textbf{if} $||\mathbf{t}|| = w_g$ 
  $w_g \gets ||\mathbf{t}||; \; S_f \gets S_f + \left\{ \mathbf{t} \right\}$  

\hspace*{0.3cm}$S_g \gets S_g + S_f$

Construct $\mathbf{H}_{r_1}$ from the rows in $S_g$ after removing duplicate rows or any cyclically shifted versions.
    \item Repeat step 2 for $\mathbf{H}_{r_1}$, replacing $\mathbf{r}_j$ with all its cyclically shifted versions $\mathbf{r}_j^{(q)}$, where $q=1,2,\ldots,N$, to yield $\mathbf{H}_{r_2}$. If the number of rows $b_r$ in $\mathbf{H}_{r_2}$ is less than $2(N-K)$, pad $2(N-K - b_r)$ rows by duplicating rows of $\mathbf{H}_{r_2}$ sequentially to increase redundancy.
    \item Apply a heuristic method, such as simulated annealing, to reduce the number of length-4 cycles in $\mathbf{H}_{r_2}$ and balance the column weight distribution, resulting in the final parity-check matrix $\mathbf{H}_s$.
\end{enumerate}

\begin{table}[htbp]
\caption{\scriptsize \uppercase{Parity-Check Matrices for Selected BCH Codes of Length 63 Before and After Optimization.}}
\label{tab:matrix-table}
\resizebox{0.49\textwidth}{!}
{%
\begin{tabular}{|c|c|c|c|c|}
\hline
Codes &
  \begin{tabular}[c]{@{}c@{}}Parity-Check \\ Matrix (Dimensions)\end{tabular} &
  \begin{tabular}[c]{@{}c@{}}Number of\\ Length-4 Cycles\end{tabular} &
  \begin{tabular}[c]{@{}c@{}}Column Weight\\ Range/Mean/Std\end{tabular} &
  \begin{tabular}[c]{@{}c@{}}Row Weight\\ Range/Mean/Std\end{tabular} \\ \hline
\multirow{2}{*}{(63,36)} & $\mathbf{H}$ (27$\times$63)      & 5909   & {[}1,13{]} / 7.7 / 3.9   & {[}18,18{]} / 18 / 0.0   \\ \cline{2-5} 
                         & $\mathbf{H}_{s}$ (183$\times$63) & 114196 & {[}44,49{]} / 46.2 / 1.3 & {[}14,16{]} / 15.9 / 0.4 \\ \hline
\multirow{2}{*}{(63,39)} & $\mathbf{H}$ (24$\times$63)      & 32625  & {[}1,18{]} / 10.7 / 5.9  & {[}28,28{]} / 28 / 0.0   \\ \cline{2-5} 
                         & $\mathbf{H}_{s}$ (41$\times$63)  & 2932   & {[}5,11{]} / 9.1 / 1.5   & {[}14,14{]} / 14 / 0.0   \\ \hline
\multirow{2}{*}{(63,45)} & $\mathbf{H}$ (18$\times$63)      & 7251   & {[}1,11{]} / 6.9 / 2.9   & {[}24,24{]} / 24 / 0.0   \\ \cline{2-5} 
                         & $\mathbf{H}_{s}$ (33$\times$63)  & 3066   & {[}7,11{]} / 8.4 / 1.0   & {[}16,16{]} / 16 / 0.0   \\ \hline
\end{tabular}
}
\end{table}

As shown in Table~\ref{tab:matrix-table}, for the (63,36) BCH code \cite{helmling19}, $\mathbf{H}_s$ has significantly more rows (increased from 27 to 183) compared to the original $\mathbf{H}$, along with a notable increase in the number of length-4 cycles. Despite both having a rank of 27, $\mathbf{H}_s$ exhibits reduced density and a smaller standard deviation ($\sigma$) in column weight distribution, both of which are favorable for NMS decoding. The adverse effects of the increased number of length-4 cycles are mitigated by the row redundancy in $\mathbf{H}_s$. Notably, it is impossible to find a set of 27 rows with weight 14 and their cyclic shifts to form a valid parity-check matrix for this code. In contrast, the derived $\mathbf{H}_s$ matrices for the BCH (63,39) and (63,45) codes consist of rows with weights 14 and 16, respectively, achieving ranks of 24 and 18. Both matrices exhibit fewer length-4 cycles than their respective original $\mathbf{H}$ matrices, despite the increased number of rows. Additionally, row weights are regular, and column weight variation is minimized in the new $\mathbf{H}_s$ matrices.
\subsection{Revised NMS}
Recognizing that permutation elements in the automorphism group of cyclic codes can be viewed as various implementations of the same transmitted codeword \cite{baldi2008iterative}, we propose to enhance the input by applying multiple permutations during each NMS iteration. The three types of permutations employed are: interleaving, which concatenates bits at even indices with those at odd indices; the Frobenius permutation defined by $2i \mod N$ for each bit index $i$; and cyclic shifting by $s \cdot d_p + d_o \mod N$, where we assume a step size of $d_p = 21$ for a code of length 63, and $s \in S_p = \{0, 1, 2\}$ with a random offset $d_o \in [0, d_p)$. Consequently, each cyclic permutation of the received sequence, combined with its Frobenius and interleaving permutations, produces three distinct sequences.

Algorithm~\ref{alg::revised_nms} outlines the complete procedures of the proposed decoder. Distinct from the others, this approach closely resembles standard NMS of maximum iterations $I$ while concealing the use of domain knowledge within the input.

\begin{algorithm}[ht]
\caption{Revised NMS Decoder}
\label{alg::revised_nms}
\begin{algorithmic}[1]
\Require
$\mathbf{y} = [y_i]_{1}^N$, $\mathbf{H}_s$, three types of permutations of the code.
\Ensure
Optimal codeword estimate $\hat{\mathbf{c}}$ for $\mathbf{y}$.
\State 
\textbf{for}{ $t = 1, 2, \ldots, I$}
\State 
\hspace{0.3cm}Incorporate randomly chosen  permutations to create a sequence block of size $3|S_p| \times N$ associated with $\mathbf{y}$, resulting in dilated inputs $\mathbf{y}_d$, $|\cdot|$ denotes set cardinality.
    \State 
\hspace{0.3cm}Execute \eqref{eq_v2c} initialized with $\mathbf{y}_d$, followed by \eqref{eq_c2v} and \eqref{eq_bit_decision} sequentially, yielding $\mathbf{L}^{(t)}$ of the same size as $\mathbf{y}_d$.
\State 
\hspace{0.3cm}Unpack $\mathbf{L}^{(t)}$ by applying inverse permutations, then average the results to obtain the sequence $l_i^{(t)}$ for $i=1,2,\ldots,N$ and its hard decision $\hat{\mathbf{c}}^{(t)}$.
\State
\hspace{0.3cm}\textbf{if}{ $\mathbf{H}_s \hat{\mathbf{c}}^{(t)^\top} = \mathbf{0}$} 
 Return $\hat{\mathbf{c}} = \hat{\mathbf{c}}^{(t)}$.
\State
\hspace{0.3cm}\textbf{else} $y_i \gets y_i + l_i^{(t)}$.
\State 
Return $\hat{\mathbf{c}} = \hat{\mathbf{c}}^{(I)}$.
\end{algorithmic}
\end{algorithm}

\subsection{DIA Model}
In prior work, we demonstrated that the DIA model, which establishes a new reliability measurement for codeword bits, can enhance an order-$p$ OSD by incorporating additional NMS failures from LDPC codes into its decodable scope. This model is also applicable to NMS failures in BCH codes. For further details on the rationale behind DIA, we refer interested readers to \cite{li2024boosting}.

\section{Simulation Results and Analysis}
\label{simulations}
The revised NMS decoder, referred to as NMS, is implemented in Python and simulated using the TensorFlow platform on Google Colab. Its sole parameter, $\alpha = 0.78$, was optimized based on the shape of the dilated input and $I = 4$. Assuming a well-trained DIA model, the hybrid decoding approach that combines NMS (with $\mathbf{H}_s$) and order-$p$ OSD (with standard $\mathbf{H}$), supported by DIA, is abbreviated as N-D-O($I, p$). To ensure accuracy, the BER curves of competitive counterparts are calibrated according to their respective sources to avoid underestimation. Notably, most of these approaches employ standard BP decoding to obtain their BER curves rather than using NMS and the FER metric.
Interested readers are encouraged to access the source code on GitHub\footnote{\url{https://github.com/lgw-frank/Short\_BCH\_Decoding\_OSD}} to derive the optimized $\mathbf{H}_s$ and replicate the decoding performance presented in this paper.

\subsubsection{Decoding Performance}
The NMS decoder is favored for its simplicity, high throughput, and independence from noise estimation. For three BCH codes of increasing rates: (63,36), (63,39), and (63,45) \cite{helmling19}, we compare the FER or BER performance of the NMS or its hybrid scheme against other decoders.

Let $F_1$ and $F_u$ denote the FER and the undetected FER of the NMS decoder, respectively, while $F_2$ represents the FER of the concatenated OSD. The comprehensive FER $F_c$ of the hybrid is calculated as follows:
\begin{equation}
    F_c = F_u + (F_1 - F_u) \cdot F_2.
\end{equation}
For longer BCH codes, $F_u$ becomes negligible due to the rarity of undetected NMS decoding errors, allowing $F_c$ to be approximated as $F_1 \cdot F_2$.

\begin{figure}[!t]
    \centering
    \subfloat[FER comparison\label{fig:fer63_36}]{
    \resizebox{0.49\linewidth}{!}{ 
	\begin{tikzpicture}[scale=0.55]
		\begin{semilogyaxis}[
			scale = 0.75,
			xlabel={$E_b/N_0$(dB)},
			ylabel={FER},
			xmin=0.0, xmax=4.3,
			ymin=1e-4, ymax=1,
			xtick={0.0,0.5,1.0,1.5,...,4.0},
			legend pos = south west,
			ymajorgrids=true,
			xmajorgrids=true,
			grid style=dashed,
			legend style={legend columns=1},
            xminorgrids=false,          
            yminorgrids=true,
			]
\addplot[
color=cyan,
mark=x,
solid,
very thin
]
coordinates {
(0.0, 0.98)
(0.5, 0.95)
(1.0, 0.85)
(1.5, 0.75)
(2.0,0.58)
(2.5,0.37)
(3.0, 0.22)
(3.5, 0.12)
(4.0, 0.062)
(4.5, 0.022)
};	
\addlegendentry{HDD\cite{helmling19}}
\addplot[
color=violet,
mark=halfcircle,
solid,
very thin
]
coordinates {
(0.0, 0.78838)
(0.5, 0.67307)
(1.0, 0.53295)
(1.5, 0.38815)
(2.0,0.2489)
(2.2,0.1983)
(2.4,0.1575)
(2.6,0.1200)
(2.8,0.0897)
(3.0, 0.06328)
(3.2, 0.04397)
(3.4, 0.03028)
(3.6,0.0193)
(3.8,0.0122)
(4.0, 0.00709)
(4.2, 0.00418)
};	
\addlegendentry{NMS(4)}

\addplot[
color=black,
mark=square,
solid,
very thin
]
coordinates {
(0.0, 0.52458)
(0.5, 0.37767)
(1.0, 0.2483)
(1.5, 0.14551)
(2.0, 0.07576)
(2.2, 0.05266)
(2.4, 0.03589)
(2.6, 0.02625)
(2.8, 0.01746)
(3.0, 0.0105)
(3.2, 0.00692)
(3.4, 0.00435)
(3.6, 0.00261)
(3.8, 0.00141)
(4.0, 0.00074)
(4.2, 0.0004)
};
\addlegendentry{ N-D-O(4,1)} 
\addplot[
color=magenta,
mark= +,
solid,
very thin
]
coordinates {
(0.0, 0.0479)
(0.5, 0.041867)
(1.0, 0.031368)
(1.5, 0.020885)
(2.0,0.013975)
(2.2,0.010533)
(2.4,0.008214)
(2.6,0.006437)
(2.8,0.0050)
(3.0, 0.003337)
(3.2, 0.002340)
(3.4, 0.001628)
(3.6, 0.001014)
(3.8, 0.000651)
(4.0, 0.000369)
(4.2,0.000212)
};	
\addlegendentry{Undetected FER}

\addplot[
color=red,
mark= triangle,
solid,
very thin
]
coordinates {
(0.0, 0.5235)
(0.5, 0.38731)
(1.0, 0.25275)
(1.5, 0.15625)
(2.0, 0.08008)
(2.5, 0.03647)
(3.0, 0.0148)
(3.5, 0.00533)
(4.0, 0.00129)
(4.5, 0.00037)
};	
\addlegendentry{OSD(1)}
\addplot[
color=blue,
mark=asterisk,
very thin
]
coordinates {
  (0.00, 4.505e-01)
  (0.50, 3.145e-01)
  (1.00, 1.739e-01)
  (1.50, 1.018e-01)
  (2.00, 4.482e-02)
  (2.50, 1.447e-02)
  (3.00, 3.916e-03)
  (3.50, 9.864e-04)
  (4.00, 1.419e-04)
  (4.50, 2.086e-05)
  
};	
\addlegendentry{ML \cite{helmling19}}
    \end{semilogyaxis}
\end{tikzpicture}
    }}
    \hspace{-0.05\linewidth}  
    \subfloat[BER comparison\label{fig:ber63_36}]{
    \resizebox{0.48\linewidth}{!}{    
 	\begin{tikzpicture}[scale=0.55]
		\begin{semilogyaxis}[
			scale = 0.75,
			xlabel={$E_b/N_0$(dB)},
			ylabel={BER},
			xmin=1.0, xmax=4.55,
			ymin=1e-6, ymax=0.12,
			xtick={0.0,0.5,1.0,1.5,2,...,4.0},
			legend pos = south west,
			ymajorgrids=true,
			xmajorgrids=true,
			grid style=dashed,
			legend style={legend columns=1},
                xminorgrids=false, 
                yminorgrids=true,
			]
\addplot[
color=blue,
mark=diamond*,
solid,
very thin
]
coordinates {
(1.5,0.09)
(2.0,0.075)
(3.0,0.04)
(4.0,0.018)
(5.0,5.1e-3)
};	
\addlegendentry{GPT(10)\cite{babalola2019generalized}}

\addplot[
color=teal,
mark=*,
very thin
]
coordinates {
(1.0,1.1e-1)
(2.0,7e-2)
(3.0,4e-2)
(4.0,1.35e-2)
(5.0,3.3e-3)
};
\addlegendentry{ BP-RNN \cite{nachmani18}} 
\addplot[
color=cyan,
mark=triangle,
solid,
very thin
]
coordinates {
(2.0,0.048)
(2.5,0.023)
(3.0,0.011)
(3.5,0.0042)
(4.0,1.3e-3)
(4.5,3.6e-4)
(5.0,1e-4)
};	
\addlegendentry{RNN-SS(RRD)\cite{lian2019learned}}

\addplot[
color=violet,
mark=halfcircle,
solid,
very thin
]
coordinates {
(0.0, 0.1255)
(0.5, 0.1017)
(1.0, 0.0768)
(1.5, 0.0534)
(2.0,0.0331)
(2.2,0.0260)
(2.4,0.0203)
(2.6,0.0153)
(2.8,0.0113)
(3.0,0.0079)
(3.2, 0.0054)
(3.4, 0.0037)
(3.6,0.0023)
(3.8,0.0015)
(4.0,0.00084)
(4.2,0.00045)
};	
\addlegendentry{NMS(4)}
\addplot[
color=orange,
mark=diamond,
very thin
]
coordinates {
(3.0,4e-3)
(3.5,1.2e-3)
(4.0,2.4e-4)
(4.5,4e-5)
};
\addlegendentry{ mRRD-RNN(5) \cite{nachmani18}} 
\addplot[
color=black,
mark=square,
solid,
very thin
]
coordinates {
(0.0, 0.122132)
(0.5, 0.0858561)
(1.0, 0.0539239)
(1.5, 0.0308432)
(2.0, 0.0171405)
(2.2, 0.0109889)
(2.4, 0.0071799)
(2.6, 0.0055853)
(2.8, 0.0033218)
(3.0, 0.0022704)
(3.2, 0.0014426)
(3.4, 0.000874)
(3.6, 0.0005428)
(3.8, 0.0002794)
(4.0, 0.0001539)
(4.2, 8.26e-05)
};	
\addlegendentry{N-D-O(4,1)}
\addplot[
color=magenta,
mark= +,
solid,
very thin
]
coordinates {
(0.0,  0.010094)
(0.5, 0.009106)
(1.0, 0.007028)
(1.5, 0.004848)
(2.0, 0.00296)
(2.2, 0.00225)
(2.4, 0.00169)
(2.6, 0.00125)
(2.8, 0.00088)
(3.0, 0.00062)
(3.2, 0.00043)
(3.4, 0.00029)
(3.6, 0.000192)
(3.8, 0.000111)
(4.0, 0.000068)
(4.2,4e-5)
};	
\addlegendentry{Undetected BER}
\addplot[
color=blue,
mark=asterisk,
solid,
very thin
]
coordinates {
(3.0, 9e-4)
(3.5, 2.1e-4)
(4.0, 3e-5)
(4.5,3.9e-6)
};	
\addlegendentry{ML\cite{nachmani18}}
    \end{semilogyaxis}
\end{tikzpicture}}
    }
    \caption{Comparison of various decoders for the BCH (63,36) code.}
    \label{fig:fer_n_ber_63_36_5}
\end{figure}
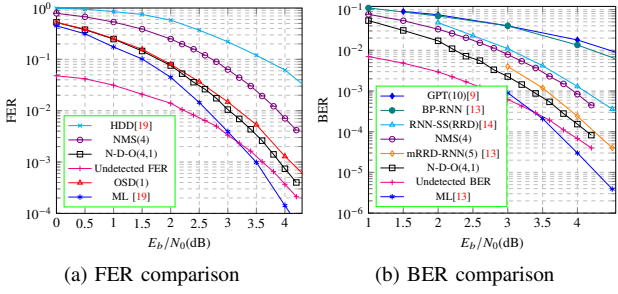

As shown in Fig.~\ref{fig:fer_n_ber_63_36_5}, in terms of FER, the proposed NMS with $I=4$ significantly outperforms hard-decision decoding (HDD), and its combination with order-1 OSD, facilitated by the DIA model, yields an additional gain of approximately 0.8 dB over NMS alone. Interestingly, the undetected FER, which measures the ratio of false positives for the NMS, intersects with the simulated ML curves at an SNR of 3.2 dB, indicating that the maxima of the ML and undetected FER curves serve as a lower bound for the hybrid approach across the entire SNR range. To the best of the authors' knowledge, this is the first report of the undetected FER (or BER) in the context of iterative decoding for short BCH codes. Regarding the BER comparison, the GPT and BP-RNN decoders significantly lag behind NMS, while the RRD adapted by RNN-SS, despite its high complexity, is slightly inferior to the NMS. Although the mRRD adapted by the RNN structure, consisting of five parallelizable sub-decoders, outperforms NMS by about 0.4 dB, it still falls short of N-D-O(4,1). Furthermore, the N-D-O(4,1) approaches undetected BER in the high SNR region.

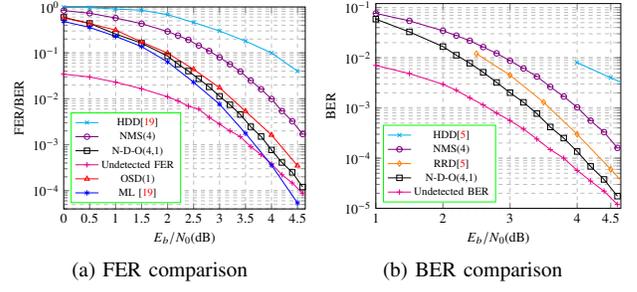
\begin{figure}[!t]
    \centering
    \subfloat[FER comparison\label{fig:fer63_39}]{
    \resizebox{0.49\linewidth}{!}{ 
    	\begin{tikzpicture}[scale=0.55]
		\begin{semilogyaxis}[
			scale = 0.75,
			xlabel={$E_b/N_0$(dB)},
			ylabel={FER/BER},
			xmin=0.0, xmax=4.65,
			ymin=4e-5, ymax=1,
			xtick={0.0,0.5,1.0,1.5,2,...,4.5},
			legend pos = south west,
			ymajorgrids=true,
			xmajorgrids=true,
			grid style=dashed,
			legend style={legend columns=1},
                xminorgrids=false, 
                yminorgrids=true,
			]
\addplot[
color=cyan,
mark=x,
solid,
very thin
]
coordinates {
(0.0, 0.99)
(0.5, 0.97)
(1.0, 0.90)
(1.5, 0.85)
(2.0,0.68)
(2.5,0.46)
(3.0, 0.3)
(3.5, 0.18)
(4.0, 0.1)
(4.5, 0.04)
};	
\addlegendentry{HDD\cite{helmling19}}
\addplot[
color=violet,
mark=halfcircle,
solid,
very thin
]
coordinates {
(0.0, 0.835)
(0.5, 0.7295)
(1.0, 0.5745)
(1.5, 0.432)
(2.0,0.29171)
(2.2,0.23802)
(2.4,0.19198)
(2.6,0.14434)
(2.8,0.11022)
(3.0, 0.0798)
(3.2, 0.05647)
(3.4, 0.03933)
(3.6, 0.0249)
(3.8, 0.01617)
(4.0, 0.00993)
(4.2, 0.00542)
(4.4, 0.00324)
(4.6, 0.0017)
};	
\addlegendentry{NMS(4)}

\addplot[
color=black,
mark=square,
]
coordinates {
(0.0, 0.59246)
(0.5, 0.43815)
(1.0, 0.257)
(1.5, 0.16267)
(2.0, 0.08521)
(2.2, 0.05691)
(2.4, 0.04017)
(2.6, 0.02719)
(2.8, 0.01817)
(3.0, 0.01133)
(3.2, 0.00741)
(3.4, 0.00455)
(3.6, 0.00246)
(3.8, 0.00151)
(4.0, 0.00077)
(4.2, 0.00041)
(4.4, 0.00025)
(4.6, 0.00012)
};
\addlegendentry{N-D-O(4,1)}
\addplot[
color=magenta,
mark= +,
solid,
very thin
]
coordinates {
(0.0, 0.0344)
(0.5, 0.0299)
(1.0, 0.023)
(1.5, 0.016544)
(2.0,0.011091)
(2.2,0.009000)
(2.4,0.006925)
(2.6,0.005971)
(2.8,0.003934)
(3.0, 0.002810)
(3.2, 0.002000)
(3.4, 0.001500)
(3.6, 0.000906)
(3.8, 0.000607)
(4.0, 0.000382)
(4.2,0.000222)
(4.4,0.000147)
(4.6,0.000087)
};	
\addlegendentry{Undetected FER}
\addplot[
color=red,
mark=triangle,
very thin
]
coordinates {
(0.0, 0.57889)
(0.5, 0.435)
(1.0, 0.31353)
(1.5, 0.17276)
(2.0, 0.09882)
(2.5, 0.04395)
(3.0, 0.01742)
(3.5, 0.00536)
(4.0, 0.00163)
(4.5, 0.00035)
};	
\addlegendentry{OSD(1)}
\addplot[
color=blue,
mark=asterisk,
very thin
]
coordinates {
  (0.00, 4.831e-01)
  (0.50, 3.584e-01)
  (1.00, 2.294e-01)
  (1.50, 1.368e-01)
  (2.00, 6.369e-02)
  (2.50, 2.296e-02)
  (3.00, 7.624e-03)
  (3.50, 1.773e-03)
  (4.00, 3.592e-04)
  (4.50, 5.403e-05)
};	
\addlegendentry{ML \cite{helmling19}}
    \end{semilogyaxis}
\end{tikzpicture}
    }}
    \hspace{-0.05\linewidth}  
    \subfloat[BER comparison\label{fig:ber63_39}]{
    \resizebox{0.48\linewidth}{!}{    
 	\begin{tikzpicture}[scale=0.55]
		\begin{semilogyaxis}[
			scale = 0.75,
			xlabel={$E_b/N_0$(dB)},
			ylabel={BER},
			xmin=1., xmax=4.65,
			ymin=1e-5, ymax=0.12,
			xtick={1.0,...,4.0,4.5},
			legend pos = south west,
			ymajorgrids=true,
			xmajorgrids=true,
			grid style=dashed,
			legend style={legend columns=1},
                xminorgrids=false, 
                yminorgrids=true,
			]
\addplot[
color=cyan,
mark=x,
thin
]
coordinates {
(4.0,8e-3)
(4.5,4e-3)
(5.0,2e-3)
};	
\addlegendentry{HDD\cite{halford2006random}}

\addplot[
color=violet,
mark=halfcircle,
solid,
very thin
]
coordinates {
(0.0,0.12047)
(0.5,0.09904)
(1.0,0.07646)
(1.5,0.05429)
(2.0,0.03447)
(2.2,0.02756)
(2.4,0.02185)
(2.6,0.01631)
(2.8,0.01223)
(3.0,0.00872)
(3.2, 0.00611)
(3.4, 0.00422)
(3.6, 0.00263)
(3.8, 0.00169)
(4.0, 0.00103)
(4.2, 0.00056)
(4.4, 0.00033)
(4.6, 0.00016)
};	
\addlegendentry{NMS(4)}
 
\addplot[
color=orange,
mark=diamond,
]
coordinates {
(2.5,1.2e-2)
(3.0,4.5e-3)
(3.5,1.3e-3)
(4.0,3e-4)
(4.5,6e-5)
(5.0,1.1e-5)
};
\addlegendentry{RRD\cite{halford2006random}}

\addplot[
color=black,
mark=square,
very thin
]
coordinates {
(0.0, 0.1163983)
(0.5, 0.0900478)
(1.0, 0.0587549)
(1.5, 0.0327839)
(2.0, 0.016677)
(2.2, 0.0112367)
(2.4, 0.007794)
(2.6, 0.0051454)
(2.8, 0.0032209)
(3.0, 0.0020139)
(3.2, 0.0012945)
(3.4, 0.0007728)
(3.6, 0.0004174)
(3.8, 0.0002509)
(4.0, 0.0001358)
(4.2, 6.85e-05)
(4.4, 3.77e-05)
(4.6, 1.75e-05)
};	
\addlegendentry{N-D-O(4,1)}

\addplot[
color=magenta,
mark=+,
very thin,
]
coordinates {
(0.0, 0.01009)
(0.5, 0.00911)
(1.0, 0.00703) 
(1.5, 0.00485)
(2.0, 0.00296) 
(2.2, 0.00225) 
(2.4, 0.00159) 
(2.6, 0.00115) 
(2.8,  0.00079) 
(3.0,  0.00056) 
(3.2, 0.00038)
(3.4, 0.00024)
(3.6, 0.000148)
(3.8, 1e-04)
(4.0, 0.000057)
(4.2,0.000035)
(4.4,0.000022)
(4.6,0.000012)
};
\addlegendentry{Undetected BER}
    \end{semilogyaxis}
\end{tikzpicture}}
    }
    \caption{Comparison of various decoders for the BCH (63,39) code.}
    \label{fig:fer_n_ber_63_39_4}
\end{figure}

For the BCH (63,39) code, as shown in Fig.~\ref{fig:fer_n_ber_63_39_4}, the NMS  clearly surpasses HDD in terms of FER, and its combination with order-1 OSD demonstrates a trend closely approaching the lower bound established by the undetected FER and ML curves. In terms of BER, although RRD outperforms NMS by about 0.35 dB, it lags behind N-D-O(4,1) by approximately 0.2 dB at a BER of $10^{-4}$. The latter approaches  undetected BER curve at SNR values beyond $E_b/N_0 = 4.5$ dB.

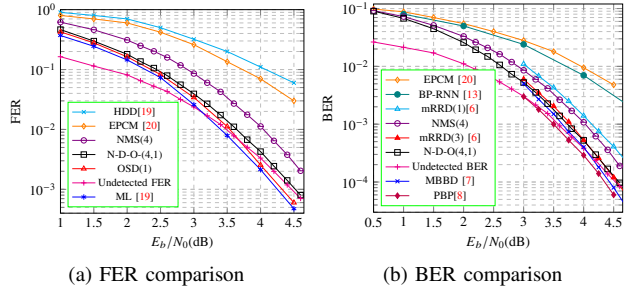
\begin{figure}[!t]
    \centering
    \subfloat[FER comparison\label{fig:fer63_45}]{
    \resizebox{0.49\linewidth}{!}{ 
	\begin{tikzpicture}[scale=0.55]
		\begin{semilogyaxis}[
			scale = 0.75,
			xlabel={$E_b/N_0$(dB)},
			ylabel={FER},
			xmin=1.0, xmax=4.65,
			ymin=4e-4, ymax=1,
			xtick={1.0,1.5,...,4.0,4.5},
			legend pos = south west,
			ymajorgrids=true,
			xmajorgrids=true,
			grid style=dashed,
            xminorgrids=false,          
            yminorgrids=true,
			]
\addplot[
color=cyan,
mark=x,
solid,
very thin
]
coordinates {
(0.0,0.98)
(0.5,0.95)
(1.0,0.9)
(1.5,0.8)
(2.0,0.7)
(2.5,0.5)
(3.0,0.32)
(3.5,0.2)
(4.0,0.11)
(4.5,0.06)
};	
\addlegendentry{HDD\cite{helmling19}}
\addplot[
color=orange,
mark=diamond,
solid,
very thin
]
coordinates {
    (0.0,0.98)
    (0.5,0.9)
    (1.0,0.8)
    (1.5,0.7)
    (2.0,0.6)
    (2.5,0.42)
    (3.0,0.26)
    (3.5,0.135)
    (4.0,0.07)
    (4.5,0.03)
};
\addlegendentry{ EPCM \cite{baldi2008iterative}} 
\addplot[
color=violet,
mark=halfcircle,
solid,
very thin
]
coordinates {
(0.0, 0.84525)
(0.5, 0.744)
(1.0, 0.61859)
(1.5, 0.46114)
(2.0,0.31048)
(2.2,0.2492)
(2.4,0.20184)
(2.6,0.15892)
(2.8,0.11785)
(3.0, 0.08572)
(3.2, 0.06015)
(3.4, 0.04245)
(3.6, 0.02701)
(3.8, 0.01761)
(4.0, 0.01124)
(4.2, 0.00647)
(4.4, 0.00378)
(4.6, 0.00204)
};	
\addlegendentry{NMS(4)}

\addplot[
color=black,
mark=square,
solid,
very thin
]
coordinates {
(0.0, 0.72706)
(0.5, 0.5899)
(1.0, 0.46454)
(1.5, 0.29848)
(2.0, 0.17992)
(2.2, 0.13633)
(2.4, 0.10585)
(2.6, 0.07915)
(2.8, 0.05583)
(3.0, 0.03927)
(3.2, 0.0268)
(3.4, 0.01738)
(3.6, 0.01107)
(3.8, 0.00675)
(4.0, 0.00427)
(4.2, 0.00244)
(4.4, 0.00141)
(4.6, 0.0008)
};
\addlegendentry{N-D-O-(4,1)}
\addplot[
color=red,
mark=triangle,
solid,
very thin
]
coordinates {
(0.0, 0.75571)
(0.5, 0.58882)
(1.0, 0.42625)
(1.5, 0.27917)
(2.0, 0.1629)
(2.5, 0.08246)
(3.0, 0.03423)
(3.5, 0.0112)
(4.0, 0.00255)
(4.5, 0.0006)
};
\addlegendentry{OSD(1)}
\addplot[
color=magenta,
mark=+,
very thin
]
coordinates {
(0.0, 0.208833)
(0.5, 0.185071)
(1.0, 0.163824)
(1.5, 0.114318)
(2.0,0.081424)
(2.2,0.065098)
(2.4,0.052520)
(2.6,0.043111)
(2.8,0.031647)
(3.0, 0.023752)
(3.2, 0.016934)
(3.4, 0.012102)
(3.6, 0.007820)
(3.8, 0.005033)
(4.0, 0.003325)
(4.2, 0.001995)
(4.4, 0.001187)
(4.6, 0.000706)
};	
\addlegendentry{Undetected FER}
\addplot[
color=blue,
mark=asterisk,
very thin
]
coordinates {
  (0.00, 6.329e-01)
  (0.50, 4.975e-01)
  (1.00, 3.704e-01)
  (1.50, 2.445e-01)
  (2.00, 1.447e-01)
  (2.50, 7.353e-02)
  (3.00, 2.595e-02)
  (3.50, 7.918e-03)
  (4.00, 2.134e-03)
  (4.5,4.751e-04)
};	

\addlegendentry{ML \cite{helmling19}}
    \end{semilogyaxis}
\end{tikzpicture}
    }}
    \hspace{-0.05\linewidth}  
    \subfloat[BER comparison\label{fig:ber63_45}]{
    \resizebox{0.49\linewidth}{!}{    
    	\begin{tikzpicture}[scale=0.55]
		\begin{semilogyaxis}[
			scale = 0.75,
			xlabel={$E_b/N_0$(dB)},
			ylabel={BER},
			xmin=0.5, xmax=4.65,
			ymin=3e-5, ymax=0.12,
			xtick={0.5,1.0,1.5,2,2.5,...,4.0,4.5},
			legend pos = south west,
   		legend style={legend columns=1},
			ymajorgrids=true,
            yminorgrids=true,
			xmajorgrids=true,
			grid style=dashed,
			]
\addplot[
color=orange,
mark=diamond,
very thin
]
coordinates {
    (0.0,0.11)
    (0.5,0.098)
    (1.0,0.088)
    (1.5,0.07)
    (2.0,0.055)
    (2.5,0.04)
    (3.0,0.028)
    (3.5,0.018)
    (4.0,0.0095)
    (4.5,0.0048)
};
\addlegendentry{EPCM \cite{baldi2008iterative}}

\addplot[
color=teal,
mark=*,
very thin
]
coordinates {
(1.0,8e-2)
(2.0,5e-2)
(3.0,2.4e-2)
(4.0,7e-3)
(5.0,1.4e-3)
};
\addlegendentry{ BP-RNN 
\cite{nachmani18}}

\addplot[
color=cyan,
mark=triangle,
very thin
]
coordinates {
(3.0, 1.1e-2)
(3.25,6.8e-3)
(3.5,4.3e-3)
(3.75,2.5e-3)
(4.0,1.4e-3)
(4.25,7.5e-4)
(4.5,4.1e-4)
(5.0,1e-4)
};
\addlegendentry{mRRD(1)\cite{dimnik2009improved}}
\addplot[
color=violet,
mark=halfcircle,
solid,
very thin
]
coordinates {
(0.0, 0.11155)
(0.5, 0.09215)
(1.0, 0.07262)
(1.5, 0.05117)
(2.0,0.03301)
(2.2,0.02613)
(2.4,0.02087)
(2.6,0.01628)
(2.8,0.01188)
(3.0,0.00856)
(3.2, 0.00596)
(3.4, 0.00416)
(3.6, 0.00264)
(3.8, 0.0017)
(4.0, 0.00108)
(4.2, 0.00062)
(4.4, 0.00036)
(4.6, 0.00019)
};	
\addlegendentry{NMS(4)}

\addplot[
color=red,
mark=triangle*,
very thin
]
coordinates {
(3.0, 6e-3)
(3.25,3.5e-3)
(3.5,2.05e-3)
(3.75,1.1e-3)
(4.0,5.3e-4)
(4.25,2.6e-4)
(4.5,1.2e-4)
(5.0,2.3e-5)
};	
\addlegendentry{mRRD(3) \cite{dimnik2009improved}}
\addplot[
color=black,
mark= square,
solid,
very thin
]
coordinates {
(0.0, 0.1188348)
(0.5, 0.0903035)
(1.0, 0.0677043)
(1.5, 0.0447321)
(2.0, 0.0258626)
(2.2, 0.0194036)
(2.4, 0.0148454)
(2.6, 0.0105967)
(2.8, 0.0073196)
(3.0, 0.0052572)
(3.2, 0.003547)
(3.4, 0.0022667)
(3.6, 0.0013916)
(3.8, 0.0008364)
(4.0, 0.0005247)
(4.2, 0.0002972)
(4.4, 0.0001685)
(4.6, 9.68e-05)
};	
\addlegendentry{N-D-O(4,1)}

\addplot[
color=magenta,
mark= +,
solid,
very thin
]
coordinates {
(0.0, 0.03206)
(0.5, 0.0264)
(1.0, 0.02137)
(1.5, 0.01708)
(2.0, 0.01108)
(2.2, 0.00885)
(2.4, 0.00699)
(2.6, 0.00539)
(2.8, 0.00406)
(3.0, 0.00303)
(3.2, 0.00215)
(3.4, 0.00153)
(3.6, 0.000936)
(3.8, 0.000597)
(4.0, 0.000385)
(4.2,0.000236)
(4.4,0.000138)
(4.6,0.000084)
};	
\addlegendentry{Undetected BER}

\addplot[
color=blue,
mark=x,
very thin
]
coordinates {
(3.0, 5e-3)
(3.25,2.7e-3)
(3.5,1.5e-3)
(3.75,7.4e-4)
(4.0,4e-4)
(4.25,1.8e-4)
(4.5,8e-5)
(5.0,1.3e-5)
};	
\addlegendentry{MBBD \cite{hehn2010multiple}}

\addplot[
color=purple,
mark= diamond*,
solid,
very thin
]
coordinates {
(3.0,3e-3)
(3.25,1.8e-3)
(3.5,1e-3)
(3.75,5.3e-4)
(4.0,2.9e-4)
(4.25,1.4e-4)
(4.5,6e-5)
};	
\addlegendentry{PBP\cite{ismail2015efficient}}
    \end{semilogyaxis}
\end{tikzpicture}}
    }  
    \caption{Comparison of various decoders for the BCH (63,45) code.}
    \label{fig:fer_n_ber_63_45_3}
\end{figure}

For the BCH (63,45) code, as shown in Fig.~\ref{fig:fer_n_ber_63_45_3}, the EPCM using a binary circulant parity-check matrix $\mathbf{H}_e$, where each row is a cyclic shift of the previous one, slightly outperforms HDD but significantly lags behind NMS. This occurs despite $\mathbf{H}_e$ having 63 rows, which is more redundant than the 33 rows in the $\mathbf{H}_s$ used by NMS, highlighting the necessity of optimizing the parity-check matrix before decoding. Again, the N-D-O(4,1) curve approaches the lower bound determined by the undetected FER and ML curves across the entire SNR range of interest. Compared to other methods, the BP-RNN and EPCM schemes are less competitive. The NMS alone slightly outperforms the mRRD with one subdecoder, while the N-D-O(4,1) performs comparably to mRRD with three subdecoders in terms of BER, remaining nearly indistinguishable from the MBBD, undetected BER, or PBP curves.

For the ablation analysis of DIA, as presented in Figs.~\ref{fig:fer_n_ber_63_36_5}-\ref{fig:fer_n_ber_63_45_3}, the hybrid N-D-O(4,1) is comparable to order-1 OSD in terms of FER across all three codes. Given that undetected FER is significant for NMS decoding of these codes, it is evident that the DIA component of the hybrid scheme sufficiently enhances the OSD to compensate for the performance loss incurred by undetected errors.
\subsubsection{Latency and Complexity Analysis}

\begin{table}[!t]
\caption{\scriptsize{\uppercase{Parameter Evaluations for Complexity Comparison of Various Decoders for BCH (63,45) Code}}}
\label{tab:complexity-table}
\resizebox{0.49\textwidth}{!}{%
\begin{tabular}{|c|ccc|c|}
\hline
\multirow{2}{*}{Decoding Schemes} &
  \multicolumn{3}{c|}{Settings} &
  \multirow{2}{*}{\begin{tabular}[c]{@{}c@{}}Complexity\\ Ratios\end{tabular}} \\ \cline{2-4}
 &
  \multicolumn{1}{c|}{($I_1,I_2,I_3$)} &
  \multicolumn{1}{c|}{\begin{tabular}[c]{@{}c@{}}\# of Permutations\\ per Iteration\end{tabular}} &
  \begin{tabular}[c]{@{}c@{}}\# of Rows in\\ Parity-Check Matrix\end{tabular} &
   \\ \hline
RRD\cite{halford2006random}               & \multicolumn{1}{c|}{(2,50,20)} & \multicolumn{1}{c|}{1}                   & 18 & 2.67   \\ \hline
mRRD(q)\cite{dimnik2009improved}           & \multicolumn{1}{c|}{(15,50,q)} & \multicolumn{1}{c|}{1}                   & 18 & q      \\ \hline
PBP\cite{ismail2015efficient}               & \multicolumn{1}{c|}{(15,50,q)} & \multicolumn{1}{c|}{1}                   & 18 & q      \\ \hline
MBBP\cite{hehn2010multiple}              & \multicolumn{1}{c|}{(66,1,3)}  & \multicolumn{1}{c|}{1 (identity)}         & 63 & 0.92   \\ \hline
BP-RNN\cite{nachmani18}            & \multicolumn{1}{c|}{$I=5$}     & \multicolumn{1}{c|}{1 (identity)}         & 18 & 0.0067 \\ \hline
NMS(4)/N-D-O(4,1) & \multicolumn{1}{c|}{$I=4$}     & \multicolumn{1}{c|}{$3|S_p| = 9$} & 33 & 0.088  \\ \hline
\end{tabular}
}
\end{table}

Table~\ref{tab:complexity-table} summarizes typical parameter evaluations from the literature for the decoders applied to the BCH (63,45) code. In the worst-case scenario, the first four decoders require $I_1 \cdot I_2 \cdot I_3$ iterations of BP per sequence. Specifically, mRRD(q) necessitates a total of $750 \cdot q$ iterations, where $q$ denotes the number of subdecoders. In contrast, PBP operates similarly to mRRD(q) but in a serial mode, allowing the discovery of any valid codeword to trigger immediate decoding termination, thus reducing complexity. RRD approximates the FER performance of mRRD(5) with 2000 iterations, as reported in Fig. 3 of \cite{ismail2015efficient}. MBBP requires only 198 iterations, significantly fewer than the previous decoders; however, its three parity-check matrices are all square, leading to higher hardware and computational complexity. Conversely, BP-RNN's complexity is equivalent to five iterative decodings, yet its FER performance remains unimpressive, as shown in Fig.~\ref{fig:fer_n_ber_63_45_3}, despite the support of structured neural networks. The proposed NMS and its hybrid N-D-O approach require only four iterations of NMS on a slightly redundant parity-check matrix. Given that order-1 OSD is sufficiently responsive, we infer that the hybrid approach incurs the least latency in decoding compared to its competitors.

Regarding computational complexity, we recognize that the complexity of a single decoder is proportional to the product of the number of rows in the parity-check matrix, the number of permutations applied per iteration, and the total iterations per sequence. Thus, we define a complexity ratio as the ratio of the complexity of various decoders to that of mRRD(1). This yields the data in the last column of Table~\ref{tab:complexity-table}, where the complexities of the DIA and OSD components in the hybrid N-D-O are safely ignored due to their relatively minor workload compared to NMS. The N-D-O scheme exhibits the lowest complexity among the decoders, except for BP-RDD. Therefore, we conclude that N-D-O provides the best trade-off between performance and complexity among existing decoders for short BCH codes. Notably, while order-1 OSD performs well for these short BCH codes with minimal computational complexity and competitive FER, its inherent serial processing nature limits its applicability in scenarios requiring high data throughput to meet stringent low-latency requirements, likely giving way to the proposed hybrid approach.

\section{Conclusion}
\label{conclusion}

In this work, we designed a scheme for classic BCH codes that systematically optimizes the parity-check matrix through a series of transformations. By aggregating various random permutations of the input per iteration, the revised NMS decoding can mitigate the detrimental effects of short cycles in the TG and significantly expedite the decoding. To address the performance loss from undetected errors that pass through the parity checks and to narrow the gap to the ML limit, the OSD approach, utilizing the reliability measurement boosting technique introduced in prior work has also demonstrated its effectiveness for BCH codes.
For longer BCH codes, we anticipate that the proposed hybrid approach will function effectively with increased iterations for NMS and higher-order variants of OSD. Furthermore, for other codes such as Reed-Solomon codes, deriving the binary images of their parity-check matrices and incorporating the specific allowed permutations into the NMS decoding design should be relatively straightforward. However, further investigation and evidence are needed to validate these intuitions in future work.

\appendices
\section{Supplementary Material} 
In this section, we elaborate on the structure of the Decoding Information Aggregation (DIA) neural network, its role within the hybrid framework of normalized min-sum (NMS) and ordered statistics decoding (OSD), and its contributions to enhancing the efficacy of OSD, as validated by simulation results for the BCH (63, 45) code. The rationale for the DIA was presented in our prior work \cite{li2024boosting}, so it is more appropriate to discuss it in the supplementary section for self-containment.

\subsection{Structure of the DIA Model}
\begin{figure}[htbp]
    \centering 
    \includegraphics[width=\linewidth]{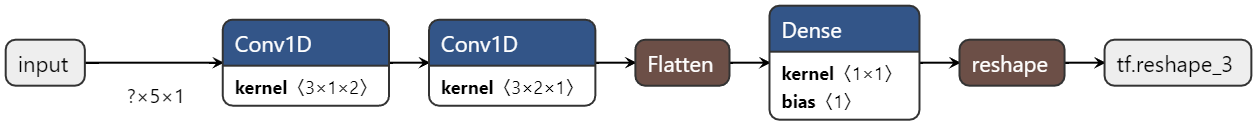} 
    \caption{DIA model to enhance bit reliability measurement for the NMS decoding failures of the BCH (63, 45) code. '?' denotes the varied batch size.}
    \label{fig:structure_DIA_63_45}
\end{figure}

As shown in Fig.~\ref{fig:structure_DIA_63_45}, the DIA model for short BCH codes consists of two off-the-shelf 'Conv1D' layers and one 'Dense' layer in the Keras package of the TensorFlow platform \cite{chollet2015keras}. It is fed with recorded trajectories of NMS decoding failures and outputs enhanced measurements.

\textbf{Input shape:} For a maximum of $I = 4$ iterations, in addition to the received sequence, a list of 5 elements is generated for each codeword bit during each NMS decoding failure after obtaining the corresponding a posteriori log-likelihood ratios (LLRs) across iterations. Typically, we choose a batch size of $b_s = 100$ for the received sequences. Therefore, there are $b_s \cdot N$ lists, each consisting of 5 elements, reshaped as input with a dimension of $6300 \times 5 \times 1$.  

\textbf{Output shape:} The output of the dense layer is $6300 \times 1$. After reshaping, the output results in a shape of $100 \times 63$, representing a batch (size of 100) of codewords composed of new measurements. 

\subsection{Framework of NMS-DIA-OSD}
\begin{figure}[htbp]
    \centering 
    \includegraphics[width=0.6\linewidth]{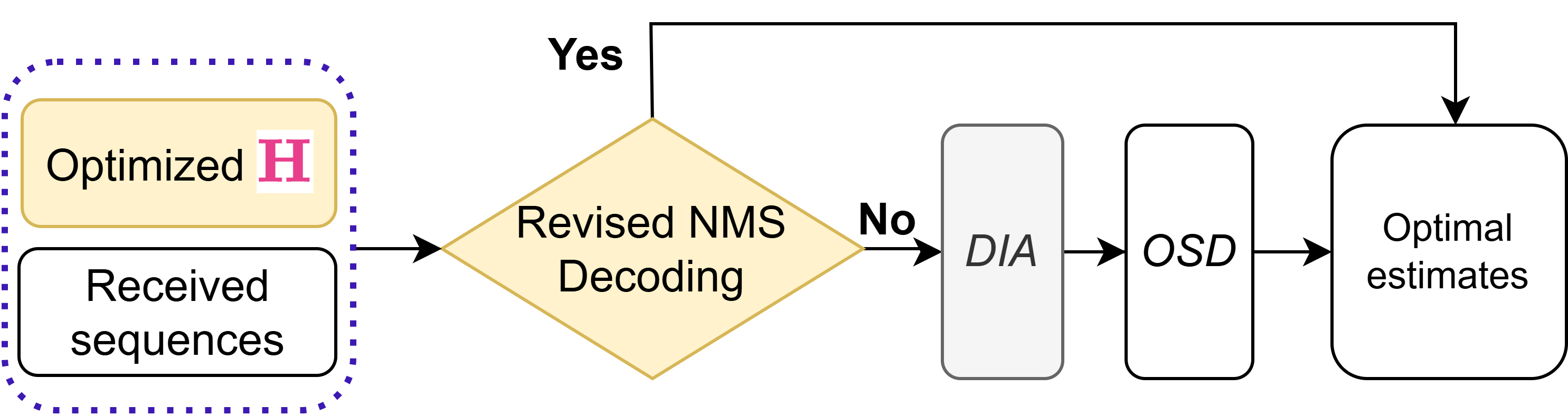} 
    \caption{The decoding framework of the BCH (63, 45) code, with the colored units being the focus of this work.}
    \label{fig:structure_NDO_63_45}
\end{figure}

As illustrated in Fig.~\ref{fig:structure_NDO_63_45}, we primarily address the following issues: how to acquire a parity-check matrix amenable to successful NMS decoding; how to revise the existing NMS to incorporate the domain knowledge of BCH codes into its iterative decoding; and how to extend the application of the DIA model for LDPC codes to a class of cyclic high-density parity-check (HDPC) codes, such as BCH codes. Specifically, the DIA model utilizes the iterative trajectories of decoding failures to enhance the measurement of codeword bits, thus boosting the functionality of OSD decoding.

In this hybrid scheme, the NMS plays a dual role. First, it handles the majority of the decoding load due to its parallelizable implementation. Second, its failures are input into the DIA model to generate better reliability measurements for the OSD. Since failed NMS occurrences are relatively rare and random across most SNR regions, they align well with the serial processing mode of OSD decoding, serving as an auxiliary role in the hybrid scheme. Therefore, we advocate for this streamlined hybrid of NMS+DIA+OSD due to its superior frame error rate (FER) capability, reduced computational complexity, and responsiveness to handling large volumes of received sequences, in addition to its invariance to noise estimation.

\subsection{Validation Results for DIA}
\begin{figure}[htbp]
    \centering
    \subfloat[The cross-entropy evaluations for the a posteriori LLRs of each NMS iteration with a maximum of $I = 4$ iterations and the enhanced measurements after the application of DIA for the BCH (63, 45) code.\label{fig:CE_DIA_63_45}]{
        \resizebox{0.42\linewidth}{!}{  \includegraphics{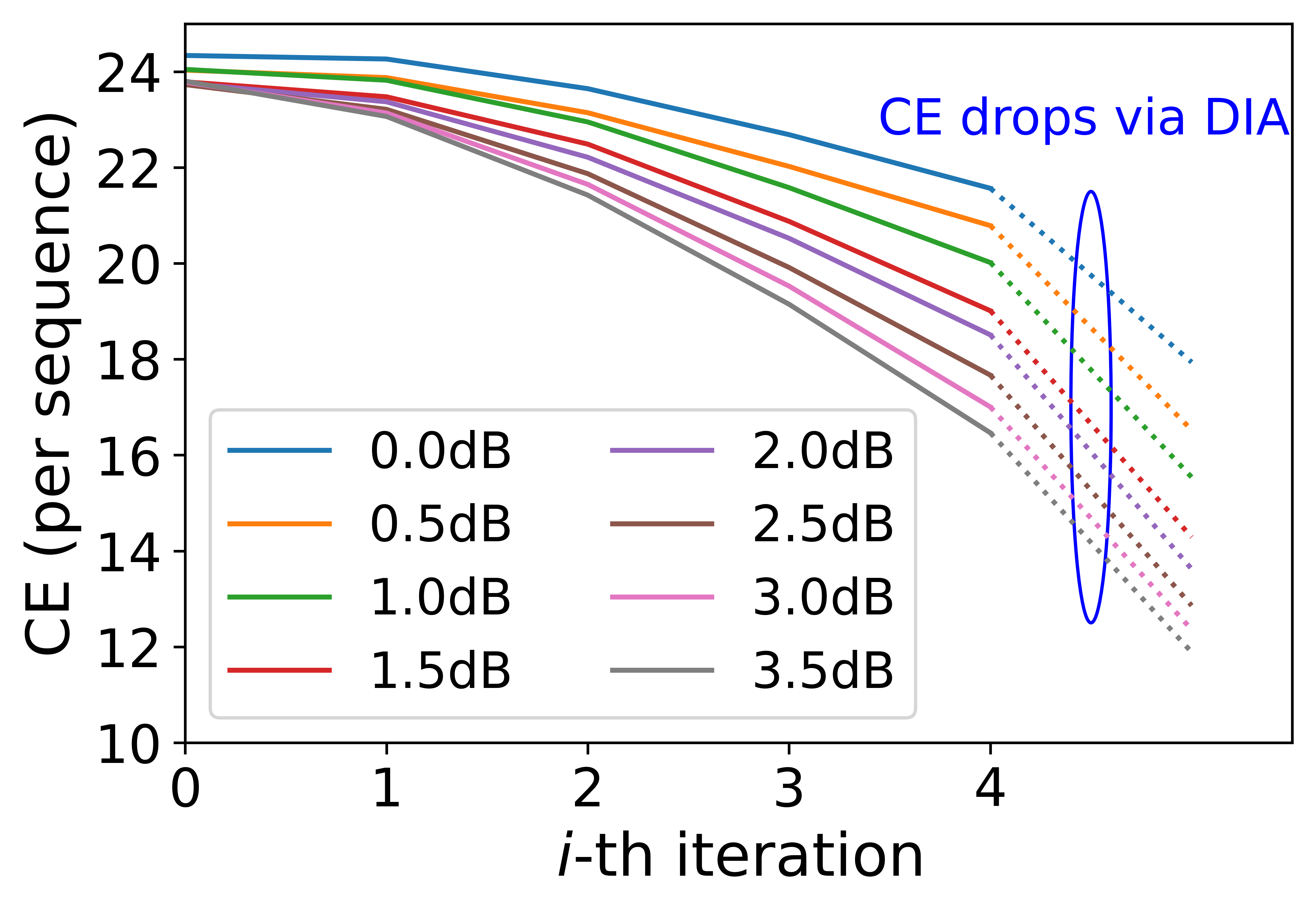} }
    }
    \hspace{0.01\linewidth}  
    \subfloat[The cumulative distribution functions (CDFs) in terms of the number of erroneous bits $\delta$ in the most reliable basis (MRB) of the BCH (63, 45) code for its NMS decoding failures at SNR = 2.6 dB.\label{fig:Errors_DIA_63_45}]{
        \resizebox{0.42\linewidth}{!}{    
        \includegraphics{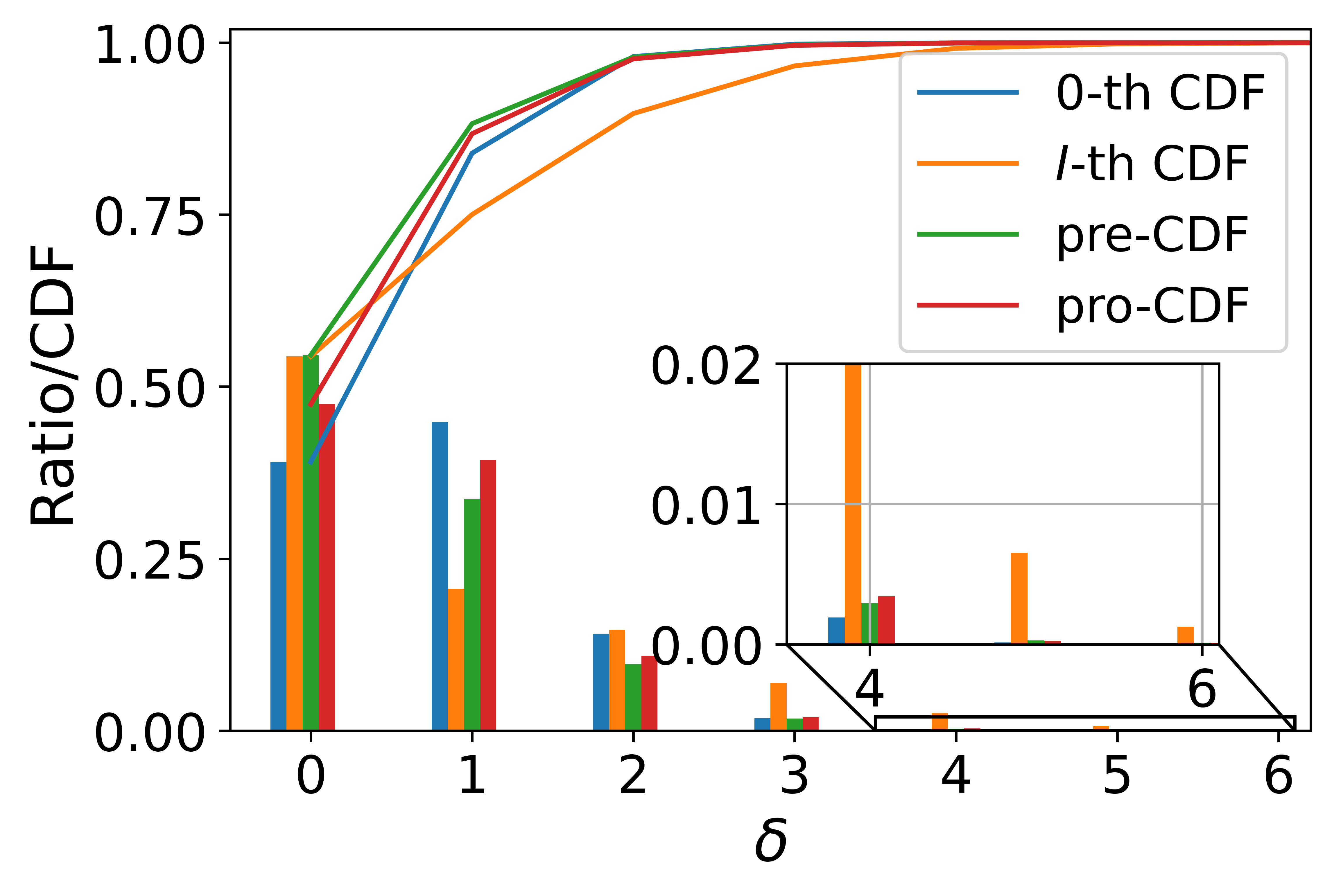} }
    }  
    \caption{Impact of DIA on the soft information and hard decisions of NMS decoding failures for the BCH (63, 45) code.}
    \label{fig:fer_n_ber_63_45_3}
\end{figure}

Regarding the cross-entropy (CE) evaluation for the a posteriori LLRs of each NMS iteration and the DIA output, Fig.~\ref{fig:CE_DIA_63_45} demonstrates that DIA can bring the soft codeword bit information closer to the ground truth compared to any NMS iterations.

On the other hand, as shown in Fig.~\ref{fig:Errors_DIA_63_45}, the benefits for the hard-decision of codeword bits in the region of the most reliable basis (MRB) for OSD are evident. Ignoring the impact of Gaussian elimination in OSD, the DIA consistently leads with its CDF ('pre-CDF') curve over the curves of the received sequence ('0-th CDF') and the a posteriori LLRs of the $I$-th iteration ('$I$-th CDF'). Notably, while the $I$-th curve shows a comparable ratio for the number of cases with no erroneous bits in the MRB, it rapidly lags behind the other measurements for cases of one bit error or two bit errors in the MRB. For codeword bits affected by Gaussian elimination in OSD, the 'pro-CDF' curve exhibits transient degradation, but this can largely be compensated by the cases of one and two errors in the MRB when compared to the 'pre-CDF' curve.

\bibliography{references/new_ref}
\end{document}